\documentclass[10pt]{article}     
\usepackage{a4wide}
\usepackage[tbtags]{amsmath}
\usepackage{accents}
\usepackage{rotating}

\DeclareMathSymbol{\vecarrow}{\mathord}{letters}{"7E}

\newlength{\lvech}
\newlength{\lvecw}

\newcommand{\lvec}[1]{\ensuremath{
\text{\settoheight{\lvech}{$\vecarrow$}\addtolength{\lvech}{-.047ex}\settowidth{\lvecw}{$\vecarrow$}$\accentset{\hspace{.47\lvecw}\begin{rotate}{180}\makebox[0pt]{\raisebox{-\lvech}[0pt][0pt]{$\vecarrow$}}\end{rotate}}{#1}$}}}

\newcommand{\sE}{{\scriptscriptstyle E}}
\renewcommand{\theequation}{\arabic {section}.\arabic{equation}}

\def\msr{{\rm I\!R}} 

\begin{document}
\makeatletter
\title{Deformation Quantization in the Teaching of Quantum
Mechanics}
\author{Allen C.
Hirshfeld\footnote{hirsh@physik.uni-dortmund.de}\, and Peter
Henselder\footnote{henselde@dilbert.physik.uni-dortmund.de}\\
Fachbereich Physik\\ Universit\"at Dortmund\\ 44221 Dortmund}

\maketitle

\begin{abstract}
We discuss the deformation quantization
approach for the teaching of quantum mechanics. This approach has
certain conceptual advantages that make its consideration
worthwhile. In particular, it sheds new light on the relation
between classical and quantum mechanics. We demonstrate
how it can be used to solve specific problems and clarify its
relation to conventional quantization and path integral
techniques. We also discuss its recent
applications in relativistic quantum field theory.
\end{abstract}

\section{Introduction}
In this article we discuss and compare three approaches to quantum
mechanics: the operator formalism, the path integral approach, and
deformation quantization. Conventional texts use the formalism in which
the observables are represented by operators in Hilbert space,
an approach that goes back to Dirac$^{\ref{bib:Dirac}}$ and
von Neumann.$^{\ref{bib:von Neumann}}$ The path integral approach
was initiated by Feynman$^{\ref{bib:Feynman}}$ and is widely used
today in research in quantum field theory. For this reason it is
also discussed in some introductory treatments, for example, in
Ref.~\cite{Sakurai},\ and in more advanced
texts.$^{\ref{bib:Peskin}}$ Some of the techniques used in
deformation quantization were introduced by the pioneers of
quantum mechanics (Wigner,$^{\ref{bib:Wigner}}$
Weyl,$^{\ref{bib:Weyl}}$ von Neumann$^{\ref{bib:von Neumann}}$),
but it was first proposed as an autonomous theory by Bayen et
al.$^{\ref{bib:Bayen}}$ in 1978. Since then, many other articles
have been written on the topic. For recent reviews see
Refs.~\cite{Sternheimer}--\cite{Zachosrev}.

Deformation quantization concentrates on the central physical
concepts of quantum theory: the algebra of observables and their
dynamical evolution.$^{\ref{bib:Haag}}$ Because it deals exclusively
with functions of phase space variables, its conceptual break with
classical mechanics is less severe than in other approaches. It
gives the correspondence principle, which played such an important
role in the historical development, a precise formulation. It is
set in the framework of Poisson manifolds, which are an important
generalization of the usual symplectic case, and which are needed
for the formulation of gauge field theories. Many of its results
can be rigorously established using adaptions of
known techniques or completely new approaches. For these reasons it
may well lead to progress in treating problems in quantum field
theory,$^{\ref{bib:Fredenhagen}}$ as well as in ordinary quantum
mechanics, where its techniques already enjoy a measure of
popularity, see for example the references in \cite{Zachosrev}.
Recently it has received an important impetus from new developments
in mathematics,$^{\ref{bib:Kontsevich}}$ recognized by Kontsevich's
Fields medal in 1998. Some believe it will supplant, or at least
complement, the other methods in quantum mechanics and quantum
field theory. As an autonomous approach to quantum theory, its
conceptual advantages recommend its use in graduate instruction.

In this article we sketch how one might go about teaching quantum
mechanics according to this approach. Background material at the
appropriate level can be found in the texts by Marsden and
Ratiu,$^{\ref{bib:Marsden}}$ Vaisman,$^{\ref{bib:Vaisman}}$ and da
Silva and Weinstein.$^{\ref{bib:Silva}}$
In Sec.~2 we briefly
review the fundamental concepts of the Hamiltonian formalism for
classical mechanics and their generalization to the setting of
Poisson manifolds. In Sec.~3 we introduce the main tool of
deformation quantization, the star product, which deforms the
commutative classical algebra of observables into the
non-commutative quantum algebra of observables. Having chosen a
quantization scheme one can proceed to calculate the physical
quantities of interest for a given system, following the method
presented in Sec.~4. These techniques are illustrated in Sec.~5 for
the case of the simple harmonic oscillator. Section~6 gives a
short review of conventional quantization, in preparation for
Sec.~7, which compares the conventional approach to deformation
quantization and to path integral methods. Finally, Sec.~8 gives an
overview of some important applications of deformation quantization
in relativistic quantum field theory. The appendix demonstrates
calculational techniques useful in this context, and suggests some
exercises that students can do to familiarize themselves with the
material.

\section{Classical Mechanics and Poisson Manifolds}
\setcounter{equation}{0}
Quantum mechanics and the canonical Hamiltonian formalism for
classical mechanics have always been closely related. When we speak
of a classical dynamical system with a finite number of degrees of
freedom, we have in mind something like an $n$-particle system,
where the particles are specified at any time by their
instantaneous positions and momenta. That is, the state of
the system is specified as a point in the $2n$-dimensional
phase space $M$. $M$ is a smooth manifold, and in canonical
coordinates a point $x$ in $M$ is written as
$ x=(q,p)= (q_1,\ldots,q_n,p_1,\ldots,p_n)$.

The observables of the system, such as the Hamilton function
for example, are smooth real-valued functions on this phase
space. Physical quantities of the system at some time, such as the
energy, are calculated by evaluating the Hamilton function at the point
in phase space $x_0=(q_0,p_0)$ that characterizes the state of the
system at this time. The mathematical expression for this operation
is
\begin{equation}
E=\!\int\! H(q,p)\,\delta^{(2)}(q-q_0,p-p_0)\,dq\,dp,
\label{evaluate}
\end{equation} where $\delta^{(2)}$ is the two-dimensional Dirac
delta function (we assume for notational simplicity a one-particle
system). The observables of the dynamical system are
functions on the phase space, the {states} of the system are
positive functionals on the observables (here the Dirac delta
functions), and we obtain the value of the observable in a
definite state by the operation shown in Eq.~(\ref{evaluate}).

In
general, functions on a manifold are multiplied by each other in
a pointwise manner, that is, given two functions $f$ and
$g$, their product $fg$ is the function
\begin{equation} (fg)(x)=f(x)g(x).
\label{pointwise}
\end{equation}
In the context of classical mechanics we say that the
observables build a commutative algebra, and we speak of the
commutative {\em classical algebra of observables}.

In Hamiltonian mechanics
there is another way to combine two functions on phase space in
such a way that a further function on phase space results, namely
by use of the Poisson bracket
\begin{equation}
\{f,g\}(q,p)=\sum_{i=1}^n \left( \frac{\partial f}{\partial
q_i}\frac{\partial g}{\partial p_i} -\frac{\partial f}{\partial
p_i}\frac{\partial g}{\partial q_i}\right)\bigg|_{q,p}.
\label{PB}
\end{equation}
Note that here the new function results not by using
the values of the functions $f$ and $g$ at the given point in phase
space, but rather by using the values of their derivatives at this
point.

Because
expressions like those in Eq.~(\ref{PB}) are used frequently, it
is advantageous to employ a compact notation. We provide the
derivatives with vector symbols, which indicate if they act on
functions to the right or to the left. For example,
\begin{equation}
f\lvec{\partial}_{q_i}g= \frac{\partial
f}{\partial q_i}g, \qquad f \vec{\partial}_{p_i} g= f \frac
{\partial g}{\partial p_i},
\end{equation}
and Eq.~(\ref{PB}) is written as
\begin{equation}
\{f,g\}=\sum_i f \left( \lvec{\partial}_{q_i}
\vec{\partial}_{p_i}-\lvec{\partial}_{p_i}
\vec{\partial}_{q_i}\right)g.
\end{equation}
 From now on we shall use the Einstein convention of
summing over repeated indices. When there is no danger of
confusion, we sometimes suppress the indices altogether. In such
cases the notation would strictly be correct only for one-particle
systems; for many-particle systems the summed indices are left
implicit. Eq.~(\ref{PB}) can then be written as
\begin{equation}
\{f,g\}= f \left( \lvec{\partial}_{q}
\vec{\partial}_{p}-\lvec{\partial}_{p} \vec{\partial}_{q}\right)g.
\end{equation} We may abbreviate our notation further by using $x$ 
to represent points of the phase space manifold,
$x=(x_1,\ldots,x_{2n})$, and introducing the
{\em Poisson tensor}
$\alpha^{ij}$, where the indices $i,j$ run from $1$ to $2n$. In canonical
coordinates $\alpha^{ij}$ is represented by the matrix
\begin{equation}
\alpha = \left( \begin{array}{cc} 0 & -I_n\\ I_n & 0 \end{array}\right),
\end{equation} where $I_n$ is the $n\times n$ identity matrix. Then
Eq.~(\ref{PB}) becomes
\begin{equation}
\{f,g\}(x)= \alpha^{ij}\,\partial_i f(x)\,\partial_j g(x),
\label{alpha}
\end{equation} where $\partial_i=\partial/\partial x_i$.

The time development of the system is given by Hamilton's equations,
which are easily expressed in terms of the Poisson brackets:
\begin{equation}
\dot{q}_i=\frac{\partial H}{\partial p_i} = \{ q_i,H\},\quad
\dot{p}_i=-\frac{\partial H}{\partial q_i}=\{p_i,H\} .
\end{equation}
For a general observable
\begin{equation}
\dot{f}=\{ f,H\}.
\end{equation}

Because $\alpha$ transforms like a tensor with respect to coordinate
transformations, Eq.~(\ref{alpha}) may also be written in
noncanonical coordinates. In this case the components of $\alpha$
need not be constants, and may depend on the point of the
manifold at which they are evaluated. But in Hamiltonian
mechanics, $\alpha$ is still required to be invertible. A manifold
equipped with a Poisson tensor of this kind is called a
symplectic manifold. In modern treatments of mechanics, such as in
Ref.~\cite{Marsden}, one uses a more general framework. The tensor
$\alpha$ is no longer required to be invertible, but it
nevertheless suffices to define Poisson brackets via
Eq.~(\ref{alpha}), and these brackets are required to have the
properties
\begin{enumerate}

\item[(i)] $\{f,g\}= -\{g,f\}$,

\item[(ii)] $\{f,gh\}=\{f,g\}h+g\{f,h\}$,

\item[(iii)] $\{f,\{g,h\}\}+\{g,\{h,f\}\}+\{h,\{f,g\}\}=0$.

\end{enumerate}
Property (i) tells us that the Poisson bracket is antisymmetric,
property (ii) is referred to as the Leibnitz rule, and property
(iii) is called the Jacobi identity. The Poisson bracket used in
Hamiltonian mechanics satisfies all these properties, but we now
abstract these properties from the concrete prescription of
Eq.~(\ref{PB}), and define a Poisson manifold
$(M,\alpha)$ as a smooth manifold $M$ equipped with a Poisson
tensor
$\alpha$, whose components are no longer necessarily constant,
such that the bracket defined by Eq.~(\ref{alpha}) has the above
properties. It turns out that such manifolds provide a better
context for treating dynamical systems with symmetries. In fact,
they are essential for treating gauge field theories, which govern
the fundamental interactions of elementary particles.

\section{Quantum Mechanics and Star Products}
\setcounter{equation}{0}
Up to now we have been considering classical mechanics. The
essential difference between classical and quantum mechanics is
Heisenberg's uncertainty relation, which implies that in
the latter, states can no longer be represented as points in
phase space. The uncertainty is a consequence of the
non-commutativity of the quantum mechanical observables. That is,
the commutative classical algebra of observables must be
replaced by a non-commutative quantum algebra of observables.

In the conventional approach to quantum mechanics this
non-commu\-tativity is implemented by representing the quantum
mechanical observables by linear operators in Hilbert space.
Physical quantities are then represented by eigenvalues of these
operators, and physical states are related to the operator
eigenfunctions. Although these entities are somehow related to
their classical counterparts, to which they are supposed to reduce
in an appropriate limit, the precise relationship has remained
obscure, one hundred years after the beginnings of quantum
mechanics. Textbooks refer to the correspondence principle, which
guided the pioneers of the subject. Attempts to give this idea a
precise formulation by postulating a specific relation between the
classical Poisson brackets of observables and the commutators of
the corresponding quantum mechanical operators, as undertaken for
example by Dirac and von Neumann, encounter insurmountable
difficulties, as pointed out by Groenewold in 1946 in an unjustly
neglected paper.$^{\ref{bib:Groenewold}}$ In the same paper
Groenewold also wrote down the first explicit representation
of a {\em star product} (see Eq.~(\ref{Moyaldef})), without
however realizing the potential of this concept for overcoming the
difficulties that he wanted to resolve.

In the deformation quantization approach there is no such break
when going from the classical system to the corresponding quantum
system; we describe the quantum system by using the same entities
that we use to describe the classical system. The observables of
the system are described by the same functions on phase space as
their classical counterparts. Uncertainty is realized by
describing physical states as distributions on phase space that
are not sharply localized, in contrast to the Dirac delta
functions which occur in the classical case. When we evaluate an
observable in some definite state according to the quantum
analogue of Eq.~(\ref{evaluate}) (see Eq.~(\ref{Expec})),
values of the observable in a
whole region contribute to the number that we obtain, which is
thus an {\em average} value of the observable in the given state.
Non-commutativity is incorporated by introducing a non-commutative
product for functions on phase space, so that we get a new
non-commutative quantum algebra of observables.

The passage from the classical algebra of observables to the quantum
algebra of observables is performed in a continuous fashion. When
mathematicians investigate a particular structure, they try to
modify it in various ways in order to see how these modifications
affect it: which properties are preserved under certain
modifications and which properties change. If the modification
changes the structure in a continuous fashion, they speak of a
deformation. All work on deformation quantization stems from
Gerstenhaber's seminal 1964 paper,$^{\ref{bib:Gerstenhaber}}$
where he introduced the concept of a star product of smooth
functions on a manifold.

For applications to quantum mechanics we consider smooth
complex-valued functions on a Poisson manifold. A star product
$f*g$ of two such functions is a new smooth function, which, in
general, is described by an infinite power series:
\begin{equation} f*g=fg + (i\hbar ) C_1(f,g) +
O(\hbar^2)=\sum\limits_{n=0}^{\infty} (i\hbar)^n C_n(f,g).
\label{star}
\end{equation}
The first term in the series is the pointwise product
given in Eq.~(\ref{pointwise}), and $(i\hbar)$ is the {\em
deformation parameter}, which we think of as varying continuously.
If $\hbar$ is identified with Planck's constant, then what varies is
really the magnitude of the action of the dynamical system considered in
units of $\hbar$: the classical limit holds for systems with large
action. In this limit, which we express here as $\hbar\rightarrow
0$, the star product reduces to the usual product. In general,
the coefficients $C_n$ will be such that the new product is
non-commutative, and we speak of the non-commutative algebra formed
from the functions with this new multiplication law as a
deformation of the original commutative algebra, which uses
pointwise multiplication of the functions.

The expressions $C_n(f,g)$ denote functions made up of the
derivatives of the functions $f$ and $g$. Examples will be given
below, as in Eq.~(\ref{C1}). It is obvious that without further
restrictions of these coefficients, the star product is too
arbitrary to be of any use. Gerstenhaber's discovery was that the
simple requirement that the new product be associative imposes
such strong requirements on the coefficients $C_n$ that they are
essentially unique in the most important cases (up to an
equivalence relation which we shall discuss below). Formally,
Gerstenhaber required that the coefficients satisfy the following
properties:

\begin{enumerate}

\item[(1)]$\sum\limits_{j+k=n}C_j(C_k
(f,g),h)=\sum\limits_{j+k=n}C_j(f,C_k(g,h))$

\item[(2)]$C_0(f,g)=fg$

\item[(3)]$C_1(f,g)-C_1(g,f)=\{f,g\}$.
\end{enumerate}
Property (1) guarantees that the star product is
associative:
$(f*g)*h=f*(g*h)$. Property (2) means that in the limit
$\hbar\rightarrow 0$ the star product $f*g$ agrees with the
pointwise product $fg$. Property (3) has at least two aspects.
Mathematically, it anchors the new product to the given structure
of the Poisson manifold. Physically, it provides the connection
between the classical and quantum behavior of the dynamical system.
Define a commutator by using the new product:
\begin{equation} [f,g]_* = f*g-g*f.
\end{equation}
Property (3) may then be written as
\begin{equation}
\lim_{\hbar\rightarrow 0}\frac{1}{i\hbar}[f,g]_*=\{f,g\}.
\label{corres}
\end{equation}
Equation~(\ref{corres}) is the correct form of the correspondence
principle. In general, the quantity on the left-hand side of
Eq.~(\ref{corres}) reduces to the Poisson bracket only in the
classical limit. The source of the mathematical difficulties that
previous attempts to formulate the correspondence principle
encountered was related to trying to enforce {\em equality}
between the Poisson bracket and the corresponding expression
involving the quantum mechanical commutator.
Equation~(\ref{corres}) shows that such a relation in
general only holds up to corrections of higher order in
$\hbar$.

For physical applications we usually require the star product to be {\em Hermitean}: $\overline{f*g}=\bar{g}*\bar{f}$, where $\bar{f}$ denotes the complex conjugate of $f$. The star products we are mainly concerned with in the following have this property.

For a given Poisson manifold it is not clear a priori if a
star product for the smooth functions on the manifold actually
exists, that is, whether it is at all possible to find coefficients
$C_n$ that satisfy the above list of properties. Even if we find
such coefficients, it it still not clear that the series they
define through Eq.~(\ref{star}) yields a smooth function.
Mathematicians have worked hard to answer these questions in the
general case.$^{\ref{bib:DeWilde}}$ For flat Euclidian spaces,
$M=\msr^{2n}$, a specific star product has long been known. In this case the
components of the Poisson tensor
$\alpha^{ij}$ can be taken to be constants. The coefficient $C_1$
can then be chosen antisymmetric, so that
\begin{equation} C_1(f,g)=\frac{1}{2}\alpha^{ij}(\partial_i f)(\partial_j
g)=\frac{1}{2}\{f,g\},
\label{C1}
\end{equation} by property (3) above. The higher order coefficients may
be obtained by exponentiation of $C_1$. This procedure yields the
{\em Moyal star product}:$^{\ref{bib:Groenewold},\ref{bib:Moyal}}$
\begin{equation} f*_{\scriptscriptstyle M} g=f\,
e^{(\frac{i\hbar}{2})\alpha^{ij}\lvec{\partial}_i\vec{\partial}_j}\,
g.
\label{Moyaldef}
\end{equation}
In canonical coordinates Eq.~(\ref{Moyaldef}) becomes
\begin{eqnarray}
(f*_{\scriptscriptstyle M}g)(q,p) & = & f(q,p)\,e^{
\frac{i\hbar}{2}( \lvec{\partial}_q \vec{\partial}_p -\lvec{\partial}_p
\vec{\partial}_q )}\, g(q,p)\label{Moyal}\\
{} & = &
\sum\limits_{m,n=0}^\infty \left( \frac{i\hbar}{2} \right)^{m+n}
\frac{(-1)^m}{m!n!}(\partial_p^m \partial_q^n f)( \partial_p^n
\partial_q^m g).
\label{expand}
\end{eqnarray}

We now come to the question of uniqueness of the star product on a
given Poisson manifold. Two star products $*$ and $*'$ are said to
be {\em $c$-equivalent} if there exists an invertible {\em
transition operator}
\begin{equation} T=1+\hbar T_1 + \cdots =\sum\limits_{n=0}^{\infty}
\hbar^n T_n,
\label{transition}
\end{equation} where the $T_n$ are differential operators, that
satisfies
\begin{equation} f*'g=T^{-1}((T f) * (T g)).
\label{equivalence}
\end{equation} It is known that for $M=\msr^{2n}$ all admissible
star products are $c$-equivalent to the Moyal product. An example
of another star product in $\msr^{2n}$ is the {\em standard star
product}, defined by
\begin{equation} f*_{\scriptscriptstyle S} g=f\,e^{i\hbar
\lvec{\partial}_q \vec{\partial}_p}\, g.
\label{standardp}
\end{equation}
The Moyal and standard star products are $c$-equivalent,
that is,
\begin{equation}
T (f*_{\scriptscriptstyle S} g)= (Tf)
*_{\scriptscriptstyle M} (Tg),
\label{onetrans}
\end{equation}
with the transition operator
\begin{equation} T=e^{-\frac{i\hbar}{2}\vec{\partial}_q \vec{\partial}_p}.
\label{standardtransition}
\end{equation} Note that the antisymmetric part of the differential
operator in the exponent of the standard
product in Eq.~(\ref{standardp}) equals that of the exponent of the
Moyal product in Eq.~(\ref{Moyal}). This equality is a
general feature of $c$-equivalent star products: it follows from
condition (3) for the star product, which both forms have to
obey. For more general manifolds the equivalence question has been
studied in Ref.~\cite{Bertelson}. The concept of $c$-equivalence
is a mathematical one ($c$ stands for
cohomology$^{\ref{bib:Gerstenhaber}}$); it does not by itself imply
any kind of physical equivalence, as we shall see below.

Before
concluding this section we present two alternative expressions for
the Moyal star product. A form that often is useful in
calculations is given by the shift formula
\begin{equation}
(f*_{\scriptscriptstyle M} g)(q,p)=f\bigl(
q+\frac{i\hbar}{2}\vec{\partial}_p ,\, p-
\frac{i\hbar}{2}\vec{\partial}_q\bigr)\,g(q,p),
\label{shift}
\end{equation} which can be obtained from the definition,
Eq.~(\ref{expand}), by repeated applications of the Taylor formula
in the form given in Eq.~(\ref{Taylor}) in the Appendix. Still
another expression for the Moyal product, important both in theory
and applications,$^{\ref{bib:von Neumann},\ref{bib:Hansen}, \ref{bib:Rieffel}}$ is a
kind of Fourier representation:
\begin{eqnarray}
(f*_{\scriptscriptstyle M} g)(q,p) & \!\!\! = & \!
\!
\frac{1}{\hbar^2 \pi^2}
\! \int\! dq_1 dq_2 dp_1 dp_2\, f(q_1,p_1)g(q_2,p_2)
  \nonumber\\
&\!\!\!\times& \!\!\! \exp\bigl[
\frac{2}{i\hbar}\bigl(
p(q_1-q_2)+q(p_2-p_1)+(q_2p_1-q_1p_2)
\bigr].
\label{Fourier}
\end{eqnarray}
A derivation of this expression is given in the
Appendix. Equation~(\ref{Fourier}) has an interesting geometrical
interpretation.$^{\ref{bib:Zachos}}$ Denote points in phase space by
vectors, for example in two dimensions
\begin{equation}
{\bf r} = \left( \begin{array}{l}
  q\\ p \end{array}\right),\ \
{\bf r}_1 = \left(
\begin{array}{l}
  q_1\\ p_1\end{array}\right),\ \
{\bf r}_2 = \left(
\begin{array}{l}
  q_2\\ p_2 \end{array}\right).
\end{equation} Now consider the triangle in phase space spanned by the
vectors ${\bf r}-{\bf r}_1$, and ${\bf r}-{\bf r}_2$. Its area
(symplectic volume) is
\begin{eqnarray}
A({\bf r}, {\bf r}_1, {\bf r}_2) &=& \frac{1}{2} (
{\bf r}- {\bf r}_1) \wedge ( {\bf r}- {\bf r}_2 ) \nonumber \\
&=&
\frac{1}{2} [ p(q_2-q_1)+q(p_1-p_2)+(q_1p_2-q_2p_1) ],
\end{eqnarray}
which is proportional to the exponent in Eq.~(\ref{Fourier}). Hence
we may rewrite Eq.~(\ref{Fourier}) as
\begin{equation} (f*g)({\bf r})=\!\int \! d{\bf r}_1
d{\bf r}_2\, f({\bf r}_1)g({\bf r}_2)\,
\exp{ \left[ \frac{4i}{\hbar}A({\bf r},{\bf r}_1,{\bf r}_2)\right]}.
\end{equation} We shall meet this equation again at the end of this
article.

\section{Deformation Quantization}
\setcounter{equation}{0}
The properties of the star product are well adapted for describing
the noncommutative quantum algebra of observables. We have already
discussed the associativity and the incorporation of the classical
and semi-classical limits. Note that the characteristic
non-locality feature of quantum mechanics is also explicit. In the
expression for the Moyal product given in Eq.~(\ref{expand})
the star product of the functions $f$ and $g$ at the point
$x$ involves not only the values of the functions $f$ and $g$ at
this point, but also all higher derivatives of these functions at
$x$. But for a smooth function, knowledge of all the derivatives at
a given point is equivalent to knowledge of the function on the
entire space. In the integral expression of Eq.~(\ref{Fourier}) we
also see that knowledge of the functions $f$ and $g$ on the whole
phase space is necessary to determine the value of the star product
at the point $x=(q,p)$. 

The $c$-equivalent star products correspond to different
quantization schemes. Having chosen a quantization scheme, the
quantities of interest for the quantum system may be calculated.
It turns out that different quantization schemes lead to different
spectra for the observables. The choice of a specific quantization
scheme can only be motivated by further physical requirements. In
the simple example we discuss below, the classical system is
completely specified by its Hamilton function. In more general
cases one may have to decide what constitutes a sufficiently large
set of {\em good} observables for a complete specification of the
system.$^{\ref{bib:Bayen}}$

A state is
characterized by its energy
$E$, the set of all possible values for the energy is called the
spectrum of the system. The states are described by
distributions on phase space called projectors. The state
corresponding to the energy $E$ is denoted by $\pi_\sE(q,p)$. These
distributions are normalized:
\begin{equation}
\frac{1}{2\pi \hbar} \!\int\!\pi_\sE(q,p)\,dq\,dp=1,
\label{norm}
\end{equation}
and idempotent:
\begin{equation}
(\pi_\sE*\pi_{\sE'})(q,p)=\delta_{E,E'}\,
\pi_\sE(q,p).
\label{idempotent}
\end{equation} The fact that the Hamilton function takes the value $E$
when the system is in the state corresponding to this energy is
expressed by the equation
\begin{equation} (H*\pi_\sE)(q,p)=E\, \pi_\sE(q,p).
\label{stargen}
\end{equation}
Equation~(\ref{stargen}) corresponds to the time-independent
Schr\"{o}dinger equation, and is sometimes called the $*$-{\em genvalue equation}. 
The spectral decomposition of the
Hamilton function is given by
\begin{equation} H(q,p)=\sum\limits_E E\,\pi_\sE(q,p),
\label{spectral}
\end{equation} where the summation sign may indicate an integration if
the spectrum is continuous. The quantum mechanical version of
Eq.~(\ref{evaluate}) is
\begin{equation}
E=\frac{1}{2\pi \hbar} \!\int \! (H*\pi_\sE)(q,p)\,dq\,dp
=
\frac{1}{2\pi \hbar}\!\int \! H(q,p)\pi_\sE(q,p)\,dq\,dp,
\label{Expec}
\end{equation} where the last expression may be obtained by using
Eq.~(\ref{Fourier}) for the star product.

The {\em time-evolution function} for a time-independent Hamilton function is denoted by ${\rm Exp}(Ht)$,
and the fact that the Hamilton function is the generator of the
time-evolution of the system is expressed by
\begin{equation} i\hbar \frac{d}{dt}{\rm Exp}(Ht)=H*{\rm Exp}(Ht).
\label{generator}
\end{equation}
This equation corresponds to the time-dependent
Schr\"{o}dinger equation. It is solved by the {\em star exponential}:
\begin{equation}
{\rm Exp}(Ht)=\sum\limits_{n=0}^{\infty}
\frac{1}{n!}\left( \frac{-it}{\hbar}\right)^n(H*)^n,
\end{equation} where $(H*)^n=\underbrace{H*H*\cdots *H}_{{\it
n}\, {\rm times}}$. Because each state of definite energy $E$ has
a time-evolution
$e^{-iEt/\hbar}$, we expect that the complete time-evolution
function may be written in the form:
\begin{equation} {\rm Exp}(Ht)=\sum\limits_E\pi_\sE\,
e^{-iEt/\hbar}.
\label{FD}
\end{equation} This expression is called the {\em Fourier-Dirichlet expansion} 
for the time-evolution function.

Questions concerning the existence and uniqueness of the star exponential
as a $C^\infty$ function and the nature of the spectrum and the
projectors again require careful mathematical analysis. The
problem of finding general conditions on the Hamilton function $H$
which ensure a reasonable physical spectrum is analogous to the problem of
showing in the conventional approach that the symmetric operator
$\hat{H}$ is self-adjoint and finding its spectral projections. Some of
these questions have been answered by Hansen.$^{\ref{bib:Hansen}}$
Others are the subject of ongoing research. But let us
not forget that quantum mechanics is a physical theory. Physicists
are usually interested in specific systems, and one can often
determine empirically if some quantity exists just by calculating
it! This will turn out to be the case in the following example.

\section{The Simple Harmonic Oscillator}
\setcounter{equation}{0}
The simple one-dimensional harmonic oscillator is characterized by
the classical Hamilton function
\begin{equation} H(q,p)= \frac{p^2}{2m} +\frac{m \omega^2}{2} q^2 .
\end{equation} In terms of the {\em holomorphic variables}
\begin{equation}
a=\sqrt{\frac{m \omega}{2}} \left(q+ i\frac{p}{m
\omega}
\right),\ \bar{a}=
\sqrt{\frac{m \omega}{2}}\left( q- i \frac{p}{m \omega}\right)
\end{equation}
the Hamilton function becomes
\begin{equation}
H= \omega a\bar{a}.
\end{equation}

Our aim is to calculate the time-evolution function. We first choose a
quantization scheme characterized by the normal star product
\begin{equation}
f*_{\scriptscriptstyle N} g = f\, e^{\hbar
\lvec{\partial}_a \vec{\partial}_{\bar{a}}}\, g.
\end{equation} We then have
\begin{equation}
\bar{a}*_{\scriptscriptstyle N} a = a \bar{a},\
a*_{\scriptscriptstyle N} \bar{a} = a\bar{a}+\hbar,
\end{equation} so that
\begin{equation}
\left[ a, \bar{a} \right]_{*_{\scriptscriptstyle N}}=\hbar.
\end{equation}
Equation~(\ref{generator}) for this case is
\begin{equation} i \hbar \frac{d}{dt} \, {\rm Exp}_N(Ht)=( H+ \hbar
\omega
\bar{a} \partial_{\bar{a}} ) \,{\rm Exp}_N(Ht),
\end{equation} with the solution
\begin{equation}
\label{last}
{\rm Exp}_N(Ht)=e^{- a \bar{a}/\hbar}\, \exp \left(
e^{-i\omega t} a \bar{a}/\hbar \right).
\end{equation}
By expanding the last exponential in Eq.~(\ref{last}), we obtain
the
Fourier-Dirichlet expansion:
\begin{equation} {\rm Exp}_N(Ht)=e^{-a
\bar{a}/\hbar}\sum\limits_{n=0}^{\infty} \frac{1}{\hbar^n n!}
\bar{a}^n a^n\, e^{-in \omega t}.
\label{FDE}
\end{equation}
If we compare coefficients in Eqs.~(\ref{FD}) and (\ref{FDE}), we
find
\begin{eqnarray}
\pi_0^{(N)} & = & e^{-a \bar{a}/\hbar},\\
\pi_n^{(N)} & = & \frac{1}{\hbar^n n!} \, \pi_0 \bar{a}^n a^n =
\frac{1}{\hbar^n n!}\, \bar{a}^n*_{\scriptscriptstyle
N}\pi_0^{(N)}*_{\scriptscriptstyle N} a^n, \label{normal}\\ E_n &
= & n\hbar \omega.
\label{false}
\end{eqnarray}
Note that the spectrum obtained in Eq.~(\ref{false})
does not include the zero-point energy. The projector onto the
ground state $\pi_0^{(N)}$ satisfies
\begin{equation} a*_{\scriptscriptstyle N} \pi_0^{(N)}=0.
\end{equation} The spectral decomposition of the Hamilton function,
Eq.~(\ref{spectral}), is in this case
\begin{equation} H = \sum\limits_{n=0}^\infty n \hbar \omega \left(
\frac{1}{\hbar^n n!} e^{- a \bar{a}/\hbar} \bar{a}^n a^n \right)
= \omega a \bar{a}.
\end{equation} The Hamilton function is of course a classical quantity;
the factor $\hbar$ in the spectrum comes from the deformation
parameter in the star product.

We now consider the Moyal quantization scheme. If we write
Eq.~(\ref{Moyal}) in terms of holomorphic coordinates, we obtain
\begin{equation} f*_{\scriptscriptstyle M} g=f\,
e^{\frac{\hbar}{2}(\lvec{\partial}_a \vec{\partial}_{\bar{a}} -
\lvec{\partial}_{\bar{a}} \vec{\partial}_a )}\, g.
\end{equation} Here we have
\begin{equation} a *_{\scriptscriptstyle M} \bar{a} = a \bar{a} +
\frac{\hbar}{2}, \qquad
\bar{a} *_{\scriptscriptstyle M} a = a \bar{a} - \frac{\hbar}{2},
\end{equation} and again
\begin{equation}
\left[ a, \bar{a} \right]_{*_{\scriptscriptstyle M}} =\hbar.
\label{comm}
\end{equation} The value of the commutator of two phase space variables
is fixed by property (3) of the star product, and cannot change
when one goes to a $c$-equivalent star product. The Moyal star
product is $c$-equivalent to the normal star product with the
transition operator
\begin{equation} T=e^ {-\frac{\hbar}{2} \vec{\partial}_a
\vec{\partial}_{\bar{a}}} .
\end{equation}
We can use this operator to transform the normal product version of
the
\mbox{$*$-genvalue} equation, Eq.~(\ref{stargen}), into the
corresponding Moyal product version according to
Eq.~(\ref{equivalence}). The result is
\begin{equation} H*_{\scriptscriptstyle M} \pi_n^{(M)}= \omega \left(
\bar{a} *_{\scriptscriptstyle M} a + \frac{\hbar}{2}
\right)*_{\scriptscriptstyle M}\pi_n^{(M)}=
\hbar \omega \left(n+\frac{1}{2}\right)\pi_n^{(M)},
\end{equation}
with 
\begin{eqnarray}
\pi_0^{(M)} & = & T\pi_0^{(N)} = 2\, e^{-2 a
\bar{a}/\hbar},\label{pi0}\\
\pi_n^{(M)} & = & T\pi_n^{(N)} = \frac{1}{\hbar^n n!} \, \bar{a}^n
*_{\scriptscriptstyle M} \pi_0^{(M)} *_{\scriptscriptstyle M} a^n
.
\label{projectors}
\end{eqnarray}
The projector onto the ground state $\pi_0^{(M)}$
satisfies
\begin{equation}
a*_{\scriptscriptstyle M} \pi_0^{(M)} = 0\,.
\end{equation}
We now have for the spectrum
\begin{equation} E_n=\left( n+ \frac{1}{2} \right) \hbar \omega,
\label{textbook}
\end{equation}
which is the textbook result. We conclude that for this
problem the Moyal quantization scheme is the correct one.

The use of the Moyal product in Eq.~(\ref{generator}) for the star
exponential of the harmonic oscillator leads to the following
differential equation:
\begin{equation} i \hbar \frac{d}{dt} {\rm Exp}_M(Ht)=\left( H -
\frac{(\hbar \omega)^2}{4} \partial_H - \frac{(\hbar
\omega)^2}{4} H
\partial_H^2\right) {\rm Exp}_M(Ht).
\end{equation} The solution is
\begin{equation} {\rm Exp}_M(Ht)= \frac{1}{\cos{\frac{\omega
t}{2}}}\exp{\left[ \left( \frac{2H}{i\hbar\omega}\right)
\tan{\frac{\omega t}{2}} \right]}.
\label{TMoyal}
\end{equation}
This expression can be brought into the form of the
Fourier-Dirichlet expansion of Eq.~(\ref{FD}) by using the
generating function for the Laguerre
polynomials:$^{\ref{bib:Morse}}$
\begin{equation}
\label{gener}
\frac{1}{1+s}\exp{\left[
\frac{zs}{1+s}\right]}=\sum\limits_{n=0}^\infty s^n (-1)^n L_n(z),
\end{equation} with $s=e^{-i\omega t}$. The projectors then become
\begin{equation}
\pi_n^{(M)}=2(-1)^n e^{-2H/\hbar\omega}
L_n\Bigl(\frac{4H}{\hbar\omega}\Bigr),
\label{MP}
\end{equation} which is equivalent to the expression already found
in Eq.~(\ref{projectors}), as shown in the Appendix.

\section{Conventional Quantization}\label{sec:conv}
\setcounter{equation}{0}
One usually finds the observables characterizing some quantum
mechanical system by starting from the corresponding classical
system, and then, either by guessing or by using some more or less
systematic method, finding the corresponding representations of the
classical quantities in the quantum system. The guiding principle
is the correspondence principle: the quantum mechanical
relations are supposed to reduce somehow to the classical
relations in an appropriate limit. Early attempts to systematize
this procedure involved finding an assignment rule $\Theta$ that
associates to each phase space function $f$ a linear operator in
Hilbert space $\hat{f}=\Theta(f)$ in such a way that in the limit
$\hbar\rightarrow 0$, the quantum mechanical equations of motion go
over to the classical equations. Such an assignment cannot be
unique, because even though an operator that is a function of
the basic operators $\hat{Q}$ and $\hat{P}$ reduces to a unique
phase space function in the limit $\hbar\rightarrow 0$, there are
many ways to assign an operator to a given phase space function,
due to the different orderings of the operators
$\hat{Q}$ and $\hat{P}$ that all reduce to the original phase space
function. Different ordering procedures correspond to different
quantization schemes. It turns out that there is no
quantization scheme for systems with observables that depend on
the coordinates or the momenta to a higher power than quadratic
which leads to a correspondence between the quantum mechanical and
the classical equations of motion, and which simultaneously
strictly maintains the Dirac-von Neumann requirement that
$(1/i\hbar)[\hat{f},\hat{g}]
\leftrightarrow\{f,g\}$.$^{\ref{bib:Groenewold}}$ Only within the
framework of deformation quantization does the correspondence
principle acquire a precise meaning.

A general
scheme for associating phase space functions and Hilbert space
operators, which includes all of the usual orderings, is given as
follows.$^{\ref{bib:Agarwal}}$ The operator
$\Theta_\lambda(f)$ corresponding to a given phase space function
$f$ is
\begin{equation}
\Theta_\lambda(f)=\!\int \!
\tilde{f}(\xi,\eta)e^{-i(\xi\hat{Q}+\eta\hat{P})}e^{\lambda(\xi,\eta)}
d\xi \, d\eta,
\label{Weyltransform}
\end{equation} where $\tilde{f}$ is the Fourier transform of $f$,
and
$(\hat{Q},\hat{P})$ are the Schr\"{o}dinger operators that correspond to
the phase-space variables $(q,p)$; $\lambda(\xi,\eta)$ is a
quadratic form:
\begin{equation}
\lambda(\xi,\eta)=\frac{\hbar}{4}(\alpha \eta^2 + \beta \xi^2 +2i\gamma
\xi\eta ).
\end{equation} Different choices for the constants
$(\alpha,\beta,\gamma)$ yield different operator ordering
schemes. The choice $\alpha=\beta=0$ is convenient when using
$(q,p)$ coordinates;
$\gamma =1$ corresponds to {\em antistandard} ordering
\begin{equation} qp\mapsto \hat{P} \hat{Q},
\end{equation}
$\gamma=-1$ to {\em standard} ordering
\begin{equation} qp\mapsto \hat{Q}\hat{P},
\end{equation} and $\gamma=0$ to the totally symmetric {\em Weyl}
ordering
\begin{equation} qp\mapsto \frac{1}{2} (\hat{Q}\hat{P}+\hat{P}\hat{Q}).
\end{equation}

In holomorphic coordinates it is convenient to take $\gamma=0$, and
$\beta=-\alpha$. Then $\alpha=-1$ corresponds to {\em antinormal ordering}
\begin{equation} a\bar{a}\mapsto \hat{a} {\hat{a}}^\dagger,
\end{equation}
$\alpha=1$ to {\em normal ordering}
\begin{equation} a\bar{a}\mapsto \hat{a}^\dagger\hat{a},
\end{equation} and $\alpha =0$ to {\em Weyl ordering}
\begin{equation} a\bar{a}\mapsto \frac{1}{2}
(\hat{a}\hat{a}^\dagger+\hat{a}^\dagger\hat{a}).
\end{equation}

The inverse procedure of finding the phase space function that
corresponds to a given operator $\hat{f}$ is, for the special case of
Weyl ordering, given by
\begin{equation} f(q,p)=\!\int \! \langle q+\frac{1}{2}\xi | \,
\hat{f}\, |q-\frac{1}{2}\xi\rangle \, e^{-i\xi p/\hbar}
\, d\xi.
\label{FT}
\end{equation} When using holonomic coordinates it is convenient to work
with the coherent states$^{\ref{bib:Glauber}}$
\begin{equation}
\hat{a}|a\rangle=a|a\rangle,\quad
\langle\bar{a}|\hat{a}^\dagger=\langle\bar{a}|\bar{a}.
\end{equation}
These states are related to the energy eigenstates of the
harmonic oscillator
\begin{equation} |n\rangle=\frac{1}{\sqrt{n!}}\,
{\hat{a}^\dagger}^n |0\rangle
\end{equation} by
\begin{equation}
|a\rangle=e^{-\frac{1}{2}a\bar{a}/\hbar}\sum\limits_{n=0}^\infty
\frac{a^n}{\sqrt{n!}}|n\rangle,
\qquad
\langle\bar{a}|=e^{-\frac{1}{2}a\bar{a}/\hbar}\sum\limits_{n=0}^\infty
\frac{\bar{a}^n}{\sqrt{n!}}\langle n|.
\end{equation} In normal ordering we obtain the phase space function
$f(a,\bar{a})$ corresponding to the operator
$\hat{f}$ by just taking the matrix element between coherent states:
\begin{equation}
f(a,\bar{a})=\langle\bar{a}|f(\hat{a},\hat{a}^\dagger)|a\rangle.
\end{equation}

\section{Quantization, Star Products and Path Integrals}
\setcounter{equation}{0}
The relation between operator algebras and star products is given
by
\begin{equation}
\Theta(f)\Theta(g)=\Theta(f*g),
\label{Hom}
\end{equation} where $\Theta$ is a linear assignment of the kind
discussed in Sec.~\ref{sec:conv}. Different assignments, which
correspond to different operator orderings, correspond to
$c$-equivalent star products. This important relation, which was
already known to Groenewold,$^{\ref{bib:Groenewold}}$ will be
proved in the Appendix. It tells us that the quantum mechanical
algebra of observables is a representation of the star product
algebra. Because in the algebraic approach to quantum theory all
the information concerning the quantum system may be extracted
from the algebra of observables,$^{\ref{bib:Haag}}$ specifying the
star product completely determines the quantum system. In particular, if the star product is Hermitean, the operator algebra is a $C^*$-algebra. In recent
work$^{\ref{bib:Bordemann}}$ methods have been developed for
constructing explicit Hilbert space representations of the deformed
star product algebra: here the algebra of observables is the primary
object and the representing Hilbert spaces are subordinate. In
this sense deformation quantization is not just an alternative
approach to quantum theory: it may be considered as a specification
of the basic quantum structure. 

In the conventional approach the
time-development of the system is characterized by the appropriate
matrix element of the time-development operator, namely the
Feynman kernel:
\begin{equation} K(q_2,t;q_1,0)=\langle q_2| e^{-i\hat{H}t/\hbar}
| q_1
\rangle ,
\end{equation} where $\hat{H}$ is the Hamilton operator.
By substituting a complete set of energy eigenstates, we obtain
an expression resembling the Fourier-Dirichlet expansion of
Sec.~4:
\begin{equation} K(q_2,t;q_1,0)=\sum\limits_{n=0}^\infty \langle q_2|n
\rangle \langle n|q_1 \rangle e^{-i E_n t/\hbar}.
\end{equation}

For the harmonic oscillator we may insert the known
eigenfunctions and eigenvalues to obtain the following expression
for
$K(q_2,t;q_1,0)$:
\begin{eqnarray}
\frac{1}{2^n n!}
\sqrt{ \frac{m\omega}{\pi\hbar}} e^{
-\frac{m\omega}{\hbar}(q_1^2+q_2^2)}
\sum\limits_{n=0}^\infty e^{ -i(n+\frac{1}{2})\omega t} H_n\left(
\sqrt{ \frac{m\omega}{\hbar}} q_1\right) H_n\left( \sqrt{
\frac{m\omega}{\hbar}} q_2\right) \nonumber \\ =\sqrt{
\frac{m\omega} {2\pi i\hbar\sin \omega t}}\exp
\left[
\frac{im\omega}{2\hbar\sin \omega t}
\left( (q_1^2+q_2^2)\cos \omega t -2 q_1 q_2
\right)
\right],
\label{kern}
\end{eqnarray}
where we have used the following expansion formula involving the
Hermite polynomials$^{\ref{bib:Morse}}$
\begin{equation}
\frac{1}{\sqrt{1-s^2}}\exp
\left[
\frac{2xys-s^2(x^2+y^2)}{1-s^2}\right]=
\sum\limits_{n=0}^\infty
\frac{s^n}{2^n n!}H_n(x)H_n(y).
\end{equation}
In order to relate this to the phase space functions
for the harmonic oscillator discussed in Sec.~4, we apply the
Fourier transform of Eq.~(\ref{FT}) to both sides of
Eq.~(\ref{kern}). We use the following relation between the Hermite
and Laguerre polynomials:
\begin{equation}
\int \! dx \left[ H_n(x-a)H_n(x+a)e^{-x^2}
\right] e^{-2ibx}=2^n\sqrt{\pi}n! e^{-b^2}L_n(2(a^2+b^2)),
\label{HermiteLaguerre}
\end{equation}
and find
\begin{equation}
\label{eq:rhs}
2(-1)^n\sum_{n=0}^\infty
e^{-i(n+\frac{1}{2})\omega t} e^{-\frac{2H}{\hbar \omega}} L_n\left(
\frac{4H}{\hbar \omega} \right)=
\frac{1}{ \cos \frac{\omega t}{2}}\exp{
\left[ \frac{2H}{i\hbar \omega}\tan \frac{\omega t}{2} \right].}
\end{equation}
The right-hand side of Eq.~(\ref{eq:rhs}) is
the expression for the Moyal star exponential for the harmonic
oscillator as given in Eq.~(\ref{TMoyal}). According
to the Fourier-Dirichlet expansion formula (\ref{FD}) the
left-hand side gives the expressions (\ref{MP}) for the
projectors. For holomorphic coordinates the calculation is even
easier:
\begin{equation}
\pi_n^{(N)}(a,\bar{a}) =
\frac{1}{\hbar^n}\langle\bar{a}|
  n\rangle \langle n |a \rangle =\frac{1}{\hbar^n n!} (\bar{a} a)^n
e^{-\bar{a}a/\hbar},
\end{equation} in agreement with Eq.~(\ref{normal}) for the normal
star product projectors.

We see that the star exponential ${\rm Exp}(Ht)$ and the projectors
$\pi_n$ are the phase space representations of the time-evolution
operator $e^{-i\hat{H}t/\hbar}$ and the projection operators
$\hat{\rho}_n=|n\rangle \langle n|$, respectively. Weyl-ordering
corresponds to the use of the Moyal star product for quantization
and normal ordering to the use of the normal star product. In the
density matrix formalism we say that the projection operator is
that of a pure state, which is characterized by the property
of being idempotent:
$\hat{\rho}_n^2=\hat{\rho}_n$ (compare Eq.~(\ref{idempotent})). The
integral of the projector over the momentum gives the probability
distribution in position space:
\begin{eqnarray}
\frac{1}{2\pi \hbar} \! \int \! \pi_n^{(M)}(q,p)dp&=&\frac{1}{2\pi
\hbar} \! \int \! \langle q+\xi/2|n\rangle \langle n|q-\xi/2\rangle
e^{-i\xi p/\hbar}d\xi dp \nonumber\\
&=& \langle q|n\rangle \langle n|q\rangle
=|\psi_n(q)|^2,
\end{eqnarray}
and the integral over the position gives the probability
distribution in momentum space:
\begin{equation}
\frac{1}{2\pi \hbar} \! \int \! \pi_n^{(M)}(q,p)dq= \langle
p|n\rangle
\langle n|p\rangle =|\tilde\psi_n(p)|^2.
\end{equation} The normalization is
\begin{equation}
\frac{1}{2\pi \hbar}\!\int \! \pi_n^{(M)}(q,p)dq dp=1,
\end{equation} which is the same as Eq.~(\ref{norm}).
Applying these relations to the ground state projector of the
harmonic oscillator, Eq.~(\ref{pi0}) shows that this is a
minimum-uncertainty state. In the classical limit
$\hbar\rightarrow 0$, it goes to a Dirac $\delta$-function. The
expectation value of the Hamiltonian operator is
\begin{equation}
\frac{1}{2\pi \hbar}\!\int\! (H*_{\scriptscriptstyle M}
\pi_n^{(M)})(q,p)dq dp=\!\int\! \langle
q|\hat{H}\hat{\rho}_n|q\rangle dq = {\rm Tr}(\hat{H}\hat{\rho}_n),
\end{equation} which should be compared to Eq.~(\ref{Expec}).

The relation (\ref{Hom}) agrees nicely with the interpretation of the
time evolution of the quantum mechanical system suggested by Feynman. In
this view the time evolution for a finite time interval is the result of
successive steps corresponding to time intervals $\Delta t=t/N$. In the
operator formalism the finite time evolution operator is thus the product
of the short-time evolution operators
\begin{equation}
\langle q_{i+1},t_{i+1}|q_i,t_i\rangle = \langle
q_{i+1}|e^{-i\hat{H}\Delta t/\hbar}|q_i\rangle.
\end{equation}
By Eq.~(\ref{Hom}) this procedure corresponds to forming
the star product of the associated phase space functions
${\rm Exp}(H\Delta t)$. In this way the star exponential which
results for the finite time evolution is the Fourier transform of
the Feynman kernel in the sense of Eq.~(\ref{FT}). Feynman's
procedure yields the path integral expression for the kernel
\begin{equation} K(q_2,t;q_1,0)=\!\int\! Dq(t)\, e^{iS \left[ q
\right] /\hbar},
\label{pathintegral}
\end{equation} where $S\left[ q \right]$ is the classical action
functional and the notation $Dq(t)$ indicates an integration over all
paths with the fixed endpoints $q_1$ and $q_2$. For the harmonic
oscillator the semi-classical approximation is exact, and hence the
path integral can be evaluated by inserting the classical solution
into the action functional. The result of this calculation agrees
with the right-hand side of Eq.~(\ref{kern}).$^{\ref{bib:Hibbs}}$
Hence, there is also a direct relationship between the path
integral and the star exponential. This relationship has been
verified directly in the general case by Sharan$^{\ref{bib:Sharan}}$ for the
coordinate representation, and by Dito$^{\ref{bib:Dito}}$ for the
holomorphic representation.

\section{Quantum Field Theory}
\setcounter{equation}{0}
A real scalar field is given in terms of the coefficients
$ a({\mathbf k}),\bar{a}({\mathbf k}) $ by
\begin{equation}
\phi(x)=\!\int\! \frac{d^3 k}{(2\pi)^\frac{3}{2} \sqrt{2\omega_{\bf k}}}\left[
a({\bf k})e^{-ikx}+\bar{a}({\bf k})e^{ikx}\right],
\end{equation} where $\hbar \omega_{\bf k}=\sqrt{ \hbar^2{k}^2+m^2}$
is the energy of a single quantum of the field. The corresponding
quantum field operator is
\begin{equation}
\Phi(x)=\!\int\! \frac{d^3 k}{(2\pi)^\frac{3}{2} \sqrt{2\omega_{\bf
k}}}\left[
\hat{a}({\bf k})e^{-ikx}+\hat{a}^\dagger ({\bf k})e^{ikx}\right],
\end{equation} where $\hat{a}({\bf k}),\hat{a}^\dagger({\bf k})$ are the
annihilation and creation operators for a quantum of the field with
momentum $\hbar{\bf k}$. The Hamiltonian is
\begin{equation}
\label{eq:above}
H=\!\int\!d^3k\,
\hbar\omega_{\bf k}\, {\hat{a}}^\dagger({\bf k})\hat{a}({\bf k}).
\end{equation}
$N({\bf k})={\hat{a}}^\dagger({\bf k}) \hat{a}({\bf k})$ is
interpreted as the number operator, and Eq.~(\ref{eq:above}) is
then just the generalization of Eq.~(\ref{false}), the expression
for the energy of the harmonic oscillator in the normal ordering
scheme, for an infinite number of degrees of freedom. Had we
chosen the Weyl ordering scheme, we would have been lead, by the
generalization of Eq.~(\ref{textbook}), to an infinite vacuum
energy. Hence requiring the vacuum energy to vanish implies
the choice of the normal ordering scheme in free field theory. In
the framework of deformation quantization this requirement
leads to the choice of the normal star product for treating free
scalar fields, as pointed out by Dito:$^{\ref{bib:Dito}}$ only for
this choice is the star product well-defined. 

Currently, in
realistic physical field theories involving interacting
relativistic fields we are limited to perturbative calculations.
The objects of interest are products of the fields. The analog of
the {\em Moyal product} of Eq.~(\ref{Moyaldef}) for systems with
an infinite number of degrees of freedom is
\begin{eqnarray}
&&\phi(x_1)\!*\!\phi(x_2)\!*\!\cdots \!*\!\phi(x_n)
\!\!\nonumber\\
&&
=\!\!\exp{\left[
\frac{1}{2}\sum\limits_{i<j}\!\int\! d^4x\, d^4y \frac{\delta}{\delta \phi_i(x)}
\Delta (x-y)\frac{\delta}{\delta \phi_j(y)}
\right]}
\phi_1(x_1),\ldots \phi_n(x_n)|_{\phi_i=\phi},
\label{analogue}
\end{eqnarray}
where the expressions $\delta/ \delta \phi(x)$
indicate functional derivatives. Here we have used the
antisymmetric Schwinger function
\begin{equation}
\Delta(x-y)=\left[ \Phi(x),\Phi(y)\right].
\end{equation} The Schwinger function is uniquely determined by
relativistic invariance and causality from the equal-time
commutator
\begin{equation}
\left. \left[
\Phi(x),\dot{\Phi}(y)\right]\right|_{x^0=y^0}=
i\hbar\delta^{(3)}({\bf x}-{\bf y}),
\end{equation}
which is the characterization of the canonical
structure in the field theoretic framework. 

The Moyal product is,
however, not the suitable star product to use in this context. In
relativistic quantum field theory it is necessary to incorporate
causality in the form advocated by Feynman: positive
frequencies propagate {\em forward} in time, whereas negative
frequencies propagate {\em backwards} in time. This property is 
achieved by using the Feynman propagator:
\begin{equation}
\Delta_F(x) =
\left\{
\begin{array}{r@{\quad}l}
\Delta^+ (x) & {\rm for}\ x^0>0\\ -\Delta^-(x) & {\rm for}\ x^0<0,\\
\end{array}
\right.
\end{equation} where $\Delta^+(x)$, $\Delta^-(x)$ are the
propagators for the positive and negative frequency components of
the field, respectively. In operator language
\begin{equation}
\Delta_F(x-y)={\cal T}(\Phi(x)\Phi(y))-{\cal N}(\Phi(x)\Phi(y)),
\label{prop}
\end{equation}
where ${\cal T}$ indicates the time-ordered product
of the fields and ${\cal N}$ the normal-ordered product. Because
the second term in Eq.~(\ref{prop}) is a normal ordered product
with vanishing vacuum expectation value, the Feynman propagator
may be simply characterized as the vacuum expectation value of the
time-ordered product of the fields. The antisymmetric part of the
positive frequency propagator is the Schwinger function:
\begin{equation}
\Delta^+(x)-\Delta^+(-x)=\Delta^+(x)+\Delta^-(x)=\Delta(x).
\end{equation} The fact that going over to a $c$-equivalent product
leaves the antisymmetric part of the differential operator in the
exponent of Eq.~(\ref{analogue}) invariant suggests 
that the use of the positive frequency propagator instead
of the Schwinger function merely involves the passage to a
$c$-equivalent star product. This is indeed easy to verify. The time-ordered product of the operators is obtained by replacing the Schwinger function $\Delta(x-y)$ in Eq. (8.4) by the c-equivalent positive frequency propagator $\Delta^+(x-y)$, restricting the time integration to $x^0>y^0$, as in Eq. (8.7), and symmetrizing the integral in the variables $x$ and $y$, which brings in the jnegative frequency propagator $\Delta^-(x-y)$ for times $x^0<y^0$. 
Then Eq.~(\ref{analogue}) becomes Wick's theorem, which is the basic
tool of relativistic perturbation theory. In operator
language$^{\ref{bib:Leschke}}$
\begin{equation} {\cal T}(\Phi(x_1),\ldots \Phi(x_n))
=\exp{\left[
\frac{1}{2}\!\int\! d^4x\, d^4y \frac{\delta}{\delta \Phi(x)}
\Delta_F (x-y)\frac{\delta}{\delta \Phi(y)}
\right]} 
{\cal N}(\Phi(x_1),\ldots \Phi(x_n)).
\end{equation}
The relation between relativistic perturbation
theory and deformation quantization has recently been discussed by
D\"{u}tsch and Fredenhagen.$^{\ref{bib:Fredenhagen}}$

Another interesting relation between deformation quantization and quantum field
theory has been uncovered by studies of the Poisson-Sigma
model.$^{\ref{bib:Strobl}}$
This model involves a set of scalar fields $X^i$ which map a
two-dimensional manifold
$\Sigma_2$ onto a Poisson space $M$, as well as generalized gauge
fields $A_i$, which are one-forms on $\Sigma_2$ mapping to
one-forms on $M$. The action is given by
\begin{equation} S_{PS}=\!\int_{\Sigma_2} \!( A_i
dX^i+\alpha^{ij}A_i A_j ),
\end{equation} where $\alpha^{ij}$ is the Poisson structure of
$M$. The remarkable formula found by Cattaneo and
Felder$^{\ref{bib:Cattaneo}}$ is
\begin{equation} (f*g)(x)=\!\int\! DX\, DA\, f(X(1))\, g(X(2))\,
e^{iS_{PS}/\hbar},
\label{expec}
\end{equation} where $f,g$ are functions on $M$, $*$ is Kontsevich's star
product,$^{\ref{bib:Kontsevich}}$ and the functional integration is
over all fields $X$ that satisfy the boundary condition
$X(\infty)= x$. Here
$\Sigma_2$ is taken to be a disc in $\msr^2$; 1, 2, and $\infty$
are three points on its circumference. By expanding the functional
integral in Eq.~(\ref{expec}) according to the usual rules of
perturbation theory, one finds that the coefficients of the powers
of $\hbar$ reproduce the graphs and weights that characterize
Kontsevich's star product. For the case in which the Poisson tensor
is invertible, we can perform the Gaussian integration in
Eq.~(\ref{expec}) involving the fields
$A_i$. The result is
\begin{equation} (f*g)(x)=\!\int\! DX\, f(X(1))\, g(X(2))
\exp{\left[ \frac{i}{\hbar}\!
\int\! \Omega_{ij}dX^i dX^j
\right]}.
\label{Sigma1}
\end{equation}
Equation~(\ref{Sigma1}) is formally similar to Eq.~(\ref{Fourier}) for the Moyal
product, to which the Kontsevich product reduces in the symplectic
case. Here
$\Omega_{ij}=(\alpha^{ij})^{-1}$ is the symplectic 2-form, and $\int
\Omega_{ij} dX^i dX^j$ is the symplectic volume of the manifold $M$. To make this relationship exact one must integrate out the gauge degrees of freedom in the functional integral in Eq. (8.13). Since the Poisson-sigma model represents a {\em topological} field theory there would remain only a {\em finite-dimensional} integral, which would coincide with the integral in Eq. (3.17). For details of this procedure see Ref. (37).

Another important application of deformation theory in the field
theoretic context involves the correct treatment of unphysical
degrees of freedom in gauge theories. Currently it is believed that
deformation theory is the best method for dealing with these
systems.$^{\ref{bib:Stasheff}}$ There are also attempts to use the
methods of deformation quantization to treat problems in string
theory$^{\ref{bib:Garcia}}$ and quantum
gravity.$^{\ref{bib:Antonsen}}$

\noindent{\large\bf Acknowledgements}

We wish to thank Dipl.\ Cand.\ Stefan Jansen from Dortmund, who
helped us with some of the calculations which appear in this paper.
P.\ H.\ acknowledges a Stipendium for Graduate Studies from the Land
Nordrhein-Westfalen. A.\ C.\ H.\ wishes to thank the organizers of
the Third International Conference on Geometry, Integrability and
Quantization, held in Varna, Bulgaria from June 14--23, 2001, for
the opportunity to present a preliminary version of this material.

\begin{appendix}
\section{Appendix: Calculational Techniques}
\renewcommand{\theequation}{\Alph{section}.\arabic{equation}}
\setcounter{equation}{0}
Questions related to deformation quantization have attracted the
interest of many prominent mathematicians, and there is an
extensive mathematical literature on the subject. We nevertheless
wish to emphasize that physics students learning quantum mechanics
should become proficient in performing calculations in the same
style that they employ in their other subjects, for example in
classical electrodynamics. That is, they should fearlessly
interchange the order of summations and integrations, naively
manipulate delta functions, etc. Physicists have usually
achieved their results by quick and ready calculations, which were
only later justified by their mathematical colleagues. In any
case, the ultimate test for physical theories is the comparison to
experiment.

To encourage students on this path, we present
in this Appendix a few typical calculations of this kind which can
be used to motivate some of the relations in the main text. After
working through these examples, students should be able to cope
with the other results discussed.

We shall first look at the important
Eq.~(\ref{Hom}) which relates star products and
operator products. Although the relation holds for any specific
ordering and its corresponding star product, we shall perform the
calculation using Weyl ordering. The product of the two operators
$\Theta(f)$ and $\Theta(g)$, which represent the phase space
functions $f$ and
$g$, respectively, is
\begin{align}
\Theta(f) \Theta(g)& =\! \int\! d\xi_1 \, d\eta_1 \, d\xi_2 \, d
\eta_2\,\tilde{f}(\xi_1,\eta_1)\tilde{g}(\xi_2,\eta_2) \nonumber \\
&\qquad
\times \exp[ -i(\xi_1\hat{Q}+\eta_1\hat{P}) ] \exp[
-i(\xi_2\hat{Q}+\eta_2\hat{P}) ] \nonumber\\
  & = \!\int\! d \xi_1 \, d \eta_1 \, d
\xi_2 \,d \eta_2\,\tilde{f}( \xi_1,\eta_1) \tilde{g}(\xi_2,\eta_2)
\nonumber \\ &\quad \times \exp \left[ -i \left(
(\xi_1+\xi_2)\hat{Q}+(\eta_1+\eta_2)\hat{P} \right) \right] \exp
\left[\frac{-i \hbar}{2} ( \xi_1 \eta_2 - \eta_1 \xi_2) \right],
\label{Homex}
\end{align} where we have used the truncated
Campbell-Baker-Hausdorff formula:
\begin{equation} e^A e^B=e^{(A+B)}e^{\frac{1}{2}[A,B]}.
\end{equation}
We expand the last exponential in Eq.~(\ref{Homex}),
make the substitution of variables
$\xi=\xi_1+\xi_2$, $ \eta=\eta_1+\eta_2$, and obtain
\begin{multline}
\label{eq:second}
\Theta(f)\Theta(g) = \!\int\! d\xi\,d\eta\,
e^{-i(\xi\hat{Q}+\eta\hat{P})} \\
\times \!\int\! d\xi_1\,d\eta_1\sum_{m,n=0}^\infty
\frac{(-1)^m}{m!n!}\left(\frac{i\hbar}{2} \right)^{m+n}
\!\! \xi_1^m
\eta_1^n \tilde{f}(\xi_1, \eta_1)(\xi- \xi_1)^n(\eta- \eta_1)^m
\tilde{g}(\xi- \xi_1, \eta-\eta_1) .
\end{multline}
The expression on the second line of Eq.~(\ref{eq:second}) is by
the Fourier convolution theorem just the Fourier transform of the
expression for the Moyal product in Eq.~(\ref{expand}). Hence
\begin{equation}
\Theta(f)\Theta(g)=\!
\int\! d\xi\,d\eta\ \widetilde{ (f*_{\scriptscriptstyle M}g)}\
e^{-i(\xi\hat{Q}+\eta\hat{P})} =\Theta(f*_{\scriptscriptstyle M}g).
\end{equation}

To get the representation for the Moyal product of
Eq.~(\ref{Fourier}), we use again the Fourier convolution theorem
and write the Fourier transforms of the functions $f$ and $g$
explicitly:
\begin{align} f*_{\scriptscriptstyle M}g&=\frac{1}{4\pi^2}\!\int\!
d\tau\,d\sigma\,d\xi\,d\eta\,dq_1\,dp_1\,dq_2\,dp_2\,e^{i\sigma
q}e^{i\tau p}
\nonumber\\ &\times
e^{\frac{i\hbar}{2}(\eta(\sigma-\xi)-\xi(\tau-\eta))} e^{-i\xi q_1-i\eta
p_1}
\,f(q_1,p_1)e^{-i(\sigma-\xi)q_2-i(\tau-\eta)p_2}g(q_2,p_2)\nonumber\\&
=\frac{1}{4\pi^2}\!\int\!
d\tau\,d\sigma\,d\xi\,d\eta\,dq_1\,dp_1\,dq_2\,dp_2
f(q_1,p_1)g(q_2,p_2)\nonumber\\ & \times\exp\left[
i\sigma(q+\frac{\hbar}{2}\eta-q_2)+i\tau(
p-\frac{\hbar}{2}\xi-p_2) -i\xi q_1-i\eta p_1+i\xi
q_2+i\eta p_2
\right]\nonumber\\ &=\frac{1}{4\pi^2}\!\int\!
d\xi\,d\eta\,dq_1\,dp_1\,dq_2\,dp_2\ f(q_1,p_1)g(q_2,p_2)\nonumber\\
&
\times\delta\bigl(-q-\frac{\hbar}{2}\eta+q_2\bigr)\delta\bigl(-p+\frac{\hbar}{2}\xi+p_2
\bigr)
\exp[-i\xi q_1-i\eta p_1+i\xi q_2+i\eta p_2].
\end{align} Now rescale the delta functions according to
$\delta(-q-\frac{\hbar}{2}\eta+q_2)=(\frac{2}{\hbar})
\delta(\eta+\frac{2}{\hbar}q-\frac{2}{\hbar}q_2)$, and
similarly for the second delta function, and perform the $\xi$ and
$\eta$ integrations. The result is Eq.~(\ref{Fourier}).

The result in Eq.~(\ref{MP}) expressing the projectors of the
harmonic oscillator in the Moyal scheme in terms of the Laguerre
polynomials may be obtained directly by calculating the expressions
  $\pi_n^{(M)}=T\pi_n^{(N)}$ given in Eq.~(\ref{projectors}). These
expressions can be written as
\begin{eqnarray*} T\pi_n^{(N)} &=&\exp \left( -\frac \hbar
2\partial_a\partial_{\bar{a}}\right) \frac
1{\hbar^nn!}\bar{a}^na^ne^{-a\bar{a}/\hbar} \\ &=&\frac
1{\hbar^nn!}\bar{a}^na^n\exp \left( -\frac \hbar 2\left(
\lvec{\partial}_a \lvec{\partial}_{\bar{a}}+
\lvec{\partial}_a\vec{\partial}_{\bar{a}}+
\lvec{\partial}_{\bar{a}}\vec{\partial}_a+
\vec{\partial}_a\vec{\partial} _{\bar{a}}\right)
\right) e^{-a\bar{a}/\hbar}.
\end{eqnarray*}
In principle, one has to include the commutator in
the Campbell-Baker-Hausdorff formula when factorizing exponential
terms, but in this case the commutator vanishes. Hence, we can
factor out the last term in the exponent and apply Eq.~(\ref{pi0})
for $\pi_0^{(N)}$ to get
\begin{eqnarray*} &&\frac 2{\hbar^nn!}\bar{a}^na^n\exp \left( -\frac
\hbar 2\left(
\lvec{\partial} _a\lvec{\partial} _{\bar{a}}+
\lvec{\partial} _a\vec{\partial} _{\bar{a}}+
\lvec{\partial} _{\bar{a}}\vec{\partial} _a\right)
\right) e^{-2a\bar{a}/\hbar} \\ &=&\frac
2{\hbar^nn!}\bar{a}^na^n\exp
\left( -\frac \hbar 2\left(
\lvec{\partial} _a\lvec{\partial} _{\bar{a}}+
\lvec{\partial} _a\vec{\partial} _{\bar{a}}\right)
\right) \exp \left( -\frac \hbar 2\left( \lvec{\partial} _{\bar{
a}}\vec{\partial} _a\right) \right) e^{-2a\bar{a}/\hbar} \\ &=&\frac
2{\hbar^nn!}\bar{a}^na^n\exp \left( -\frac \hbar 2\left(
\lvec{\partial} _a\lvec{\partial} _{\bar{a}}+
\lvec{\partial} _a\vec{\partial} _{\bar{a}}\right)
\right) \exp \left( \lvec{\partial} _{\bar{a}}\bar{a}
\right) e^{-2a\bar{a}/\hbar}.
\end{eqnarray*}
Because the commutator $\bigl[ -\frac \hbar 2(
\lvec{\partial} _a
\lvec{\partial} _{\bar{a}}+\lvec{\partial} _a
\vec{\partial} _{\bar{a}}) ,\lvec{\partial} _{
\bar{a}}\bar{a}\bigr]$ vanishes, we can exchange the order of the
two exponentials in the last equation and then carry out the
operations indicated by the first exponential:
\begin{eqnarray*} &&\frac 2{\hbar^nn!}\bar{a}^na^n\exp \left(
\lvec{\partial} _{
\bar{a}}\bar{a}\right) \exp \left( -\frac \hbar 2\left(
\lvec{\partial} _a\lvec{\partial} _{\bar{a}}+
\lvec{\partial} _a\vec{\partial} _{\bar{a}}\right)
\right) e^{-2a\bar{a}/\hbar} \\ &=&\frac 2{\hbar^nn!}\left(
\bar{a}+\bar{a}\right)^na^n\exp
\left( -\frac \hbar 2\left( \lvec{\partial} _a\lvec{
\partial} _{\bar{a}}+\lvec{\partial} _a\vec{
\partial} _{\bar{a}}\right) \right) e^{-2a\bar{a}/\hbar} \\ &=&\frac
2{\hbar^nn!}2^n\bar{a}^na^n\exp \left( -\frac \hbar 2\left(
\lvec{\partial} _a\lvec{\partial} _{\bar{a}}\right)
\right) \exp \left( \lvec{\partial} _aa\right) e^{-2a\bar{a} /\hbar}.
\end{eqnarray*} Here we have used the Taylor formula in the form
\begin{equation} f(x+a)=e^{a\partial_x}f(x).
\label{Taylor}
\end{equation} The first exponential can now be expanded:
\begin{eqnarray*} &&\frac 2{\hbar^nn!}2^n\left( \sum_{k=0}^\infty
\frac 1{k!}\left( -\frac \hbar 2\right)^k\left( \partial
_a^ka^n\right) \left(
\partial _{\bar{a}}^k\bar{a}^n\right) \right) \exp \left(
\lvec{\partial} _a a\right) e^{-2a\bar{a}/\hbar} \\ &=&2\left(
\sum_{k=0}^n\frac{(-1)^k}{k!}\left( \frac 2\hbar \right)^{n-k}
\frac{n!}{(n-k)!(n-k)!}a^{n-k}\bar{a}^{n-k}\right) \exp \left(
\lvec{\partial}_a a\right) e^{-2a\bar{a}/\hbar} \\ &=&(-1)^n2L_n\left(
2 a\bar{a}/\hbar \right) \exp \left(
\lvec{\partial} _a a\right) e^{-2a\bar{a}/\hbar} \\ &=&(-1)^n2L_n\left(
4 a\bar{a}/\hbar \right) e^{-2a\bar{a} /\hbar},
\end{eqnarray*} where we used the definition of the Laguerre polynomials:$^{\ref{bib:Morse}}$
\begin{equation} L_n(x)=\sum_{m=0}^n (-1)^m\frac{n!}{(n-m)!m!m!}x^m.
\end{equation}

Using similar methods students should now be able to do the following
problems.
\begin{enumerate}

\item Show that Eq.~(\ref{standardtransition}) gives the transition
operator from the standard star product to the Moyal product, that
is, that Eq.~(\ref{onetrans}) is satisfied.

\item Repeat the calculation above for the projectors $\pi_n^{(M)}$ to
obtain the form of the projector $\pi_0^{(M)}$ given in
Eq.~(\ref{pi0}).

\item Verify that the projectors $\pi_0^{(N)}$ and $\pi_0^{(M)}$
satisfy the idempotence property of Eq.~(\ref{idempotent}). You may
find the Fourier form of the Moyal product, Eq.~(\ref{Fourier}),
convenient for this purpose.

\item Show that the prescriptions for passing from a phase-space
function to its corresponding Weyl-ordered operator and back, given
in Eqs.~(\ref{Weyltransform}) and (\ref{FT}), are really inverse to
each other.

\item Establish the relation between the Laguerre and Hermite
polynomials given in Eq.~(\ref{HermiteLaguerre}).

\item Perform the path integral indicated in
Eq.~(\ref{pathintegral}) for the harmonic oscillator in order to
obtain the expression (\ref{kern}) for the Feynman kernel.

\item Perform the Gaussian integration which reduces
Eq.~(\ref{expec}) to Eq.~(\ref{Sigma1}) when the Poisson tensor is
invertible.

\end{enumerate}

\end{appendix}

\end{document}